\documentclass[runningheads]{llncs}

\usepackage[T1]{fontenc}
\usepackage[utf8]{inputenc}
\usepackage{amsmath,amsfonts,amssymb}
\usepackage{stmaryrd} 
\usepackage{xspace}
\usepackage{color}
\usepackage{listings}
\usepackage{newlfont}
\usepackage{ulem} 
\usepackage{program}
\usepackage{multicol}
\usepackage{graphicx}
\usepackage{comment}
\usepackage{url}
\usepackage[noend]{algorithmic}
\usepackage{algorithm}
\usepackage{float}
\usepackage[table]{xcolor}

\usepackage{framed}
\usepackage{array}

\usepackage{lipsum}

\definecolor{lg}{rgb}{0.9, 0.9, 0.9}

\usepackage{color,listings}

\lstdefinelanguage{lustre}
    { morekeywords={
        imported, node, 
        bool, int, float,
        initial, let, tel, until, unless, type, var, when, whennot,
        match, if, then, else, state, do, done, resume, restart, returns, merge,
        pre, current, last, map, red, fill, default,
        fby, automaton, tail, implies, pre, assert, rule, query, var, declare, real,
        PROPERTY, or, and, requires, ensures, observer, rule, query, Bool, Int, ite, not},
      morecomment=[l][\color{red}]{--},
      morecomment=[s][\color{blue}]{(*}{*)},
      morecomment=[l][\color{black}]{asserts}
    }[keywords,comments]

\lstnewenvironment{lustre}
  {\lstset{language=lustre}}{}

\begin{document}
%
\title{Verifying the Safety of a Flight-Critical System}

\author{Guillaume Brat\inst{1} \and
David Bushnell\inst{2} \and
Misty Davies\inst{3} \and \\
Dimitra Giannakopoulou\inst{3} \and
Falk Howar\inst{4} \thanks{F. Howar did this work while at Carnegie Mellon University.}
\and Temesghen Kahsai\inst{1}} 
\authorrunning{Brat et. al.}
\institute{Carnegie Mellon University \and AerospaceComputing \and NASA Ames \and IPSSE, TU Clausthal}

\maketitle

\begin{abstract}
This paper describes our work on demonstrating verification 
technologies on a flight-critical system 
of realistic functionality, size, and complexity.
Our work targeted a commercial aircraft control system 
named Transport Class Model (TCM),
and involved several stages:
formalizing and disambiguating requirements in collaboration
with domain experts; processing models for their use by formal verification tools;
applying compositional techniques at the architectural and component
level to scale verification.
Performed in the context of a major NASA milestone, this study 
of formal verification in practice is
one of 
the most challenging that our group has performed, and it took several 
person months to complete it. 
This paper describes the methodology that we followed and the 
lessons that we learned.

\end{abstract}

\section{Introduction}
\label{sec:introduction}

This paper demonstrates the use of 
verification approaches on a 
safety-critical system of realistic functionality, size, 
and complexity. The work addresses a
major milestone of
the NASA Aviation Safety program and 
was performed over several months by a team involving
four verification experts, one senior software engineer,  
and an aerospace engineer.

The target of our study is a Simulink model 
of a twin-engine aircraft simulation named 
Transport Class Model (TCM). The TCM was selected 
for a number of reasons. First, it
is unclassified and can therefore be shared outside 
of NASA. This is important to us because we would like 
the community to benefit from our experience and to be 
able to use this as a common benchmark on which 
additional verification technologies can be applied. 
Second, the system was developed independently by 
a different NASA center and therefore we had no prior
knowledge of its potential errors, or its design. 
The setting was therefore similar to one in which a 
safety-critical system is handed to verification experts for 
analysis and certification, where the experts were not involved 
in the system design.

As the TCM does not come with requirements, 
we used several sources such as pilot training 
manuals and the Federal
Aviation Regulations for commercial aircraft, 
to collect relevant requirements for our study. 
A significant amount of 
our work involved formalizing and disambiguating 
requirements in collaboration
with domain experts. 
The resulting requirements constitute 
verification properties, which we encoded as 
synchronous observers in Simulink.

The Simulink models that describe the system
had to be processed in order to be usable by the verification
tools utilized in this study. 
We used SMT-based model checking, and in particular the
PKind~\cite{KahTin-PDMC-11} to verify the properties. 
The Simulink models including the synchronous observers 
were automatically translated into the synchronous dataflow language 
Lustre~\cite{lustre1}, to be processed by PKind.

A major goal of this work was to experiment with 
compositional techniques to enable the scalable use of formal
methods for systems of realistic size. 
Compositional verification constructs a verification argument for 
a complex system by composing
simpler verification results at the level of the
system components. There are several
well-known advantages to taking such 
an approach. Scalability of verification
is a major driver. Through the decomposition of
system-level requirements into component-level ones
at design time, it is easier to assign clear responsibilities to the
developer of each component. Finally, as components of a system 
change or evolve, compositional verification enables
the reuse of verification results of unchanged 
components. 

The work described in this paper is one of the most challenging 
verification exercises
that our team has performed. As such, it forced us to
define a high-level methodology for the verification
of flight-critical systems, and has enabled us to 
comment on advantages and limitations of verification
techniques and tools in handling such systems. 
We found that close collaboration between
verification and domain experts is required to formalize requirements
and, in particular, assumptions about the physical system, without which 
verification would fail. Compositional verification is key
in constructing scalable and meaningful proofs for complex systems
in the aeronautics domain. 

The remainder of the paper is organized as follows. 
Section~\ref{sec:TCM} describes the TCM, while
section~\ref{sec:requirements} discusses the process through
which we obtained requirements and disambiguated and formalized
them for verification. The verification effort itself is presented in
section~\ref{sec:verification}. The experience and lessons learned from this
substantial effort are discussed in section~\ref{sec:lessons}, with section~\ref{sec:related} placing this effort in the context of related work and 
section~\ref{sec:conclusion} concluding the paper and discussing future work.

\section{The Transport Class Model}
\label{sec:TCM}

Our target system is derived from NASA Langley's Transport Class Model (TCM)~\cite{Hueschen2011},
a simulator of a mid-size (approximately 250,000 lb.), twin-engine, commercial transport-class aircraft.
The TCM is not intended as a high-fidelity simulation of any particular transport aircraft. Rather, it 
is meant to be representative of the types of non-linear behaviors of this class of aircraft.


The TCM includes models for the avionics (with transport delay), actuators, engines, landing gear, nonlinear aerodynamics, sensors (including noise), aircraft parameters, 
equations of motion, and gravity. It is primarily implemented in Simulink, consisting of 
approximately 5700 Simulink blocks. The system also includes several thousands of lines of C/C++ code in libraries, primarily used for the engines and the nonlinear aerodynamics models.
Our work studies the guidance and controls models and their properties, within the overall context of the TCM system, and these are implemented entirely in Simulink.

\begin{figure}[t]
\centering
\includegraphics[height=2.2in]{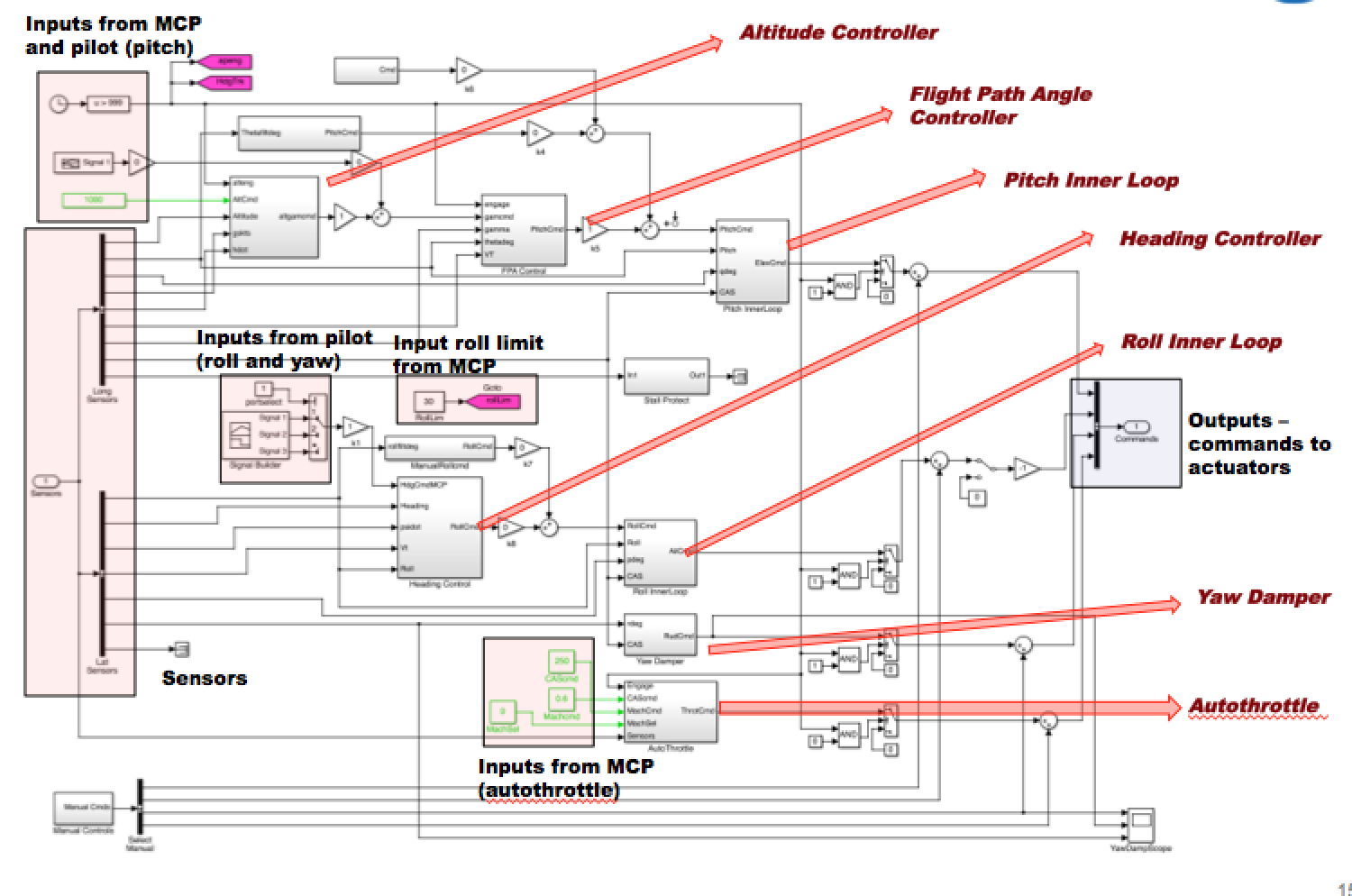}
\caption{The Simulink guidance and controls system for the TCM.}
\label{fig:Controls}
\end{figure}

\figurename~\ref{fig:Controls} depicts the top-level controls Simulink (data-flow) diagram.
An aircraft can be controlled either manually (using a control stick and pedals), or 
through the mode control panel (MCP) of the flight computer (autopilot).
In the diagram, pink-shaded boxes with dark outlines (to the left) highlight the inputs from the pilot and the MCP.
Red arrows identify the various controls subsystems (each subsystem may itself 
be a complex collection of subsystems), explained below.
The TCM contains inner loop proportional-integral (PI) controllers for all three 
angular axes of motion (roll, pitch, and yaw).
These inner loop controllers function regardless of whether the pilot flies manually 
or the autopilot issues commands.
Additionally, the TCM's autopilot can control altitude (either by flying to a directed altitude, or holding a current altitude), can reach and maintain a desired flight path angle (FPA), can reach and maintain a desired heading, and can control the airplane's speed.
Finally, the blue-shaded box with the dark outline (to the left) in the diagram shows the collected outputs---the commands to the actuators.

\section{Requirements: Elicitation and Formalization}
\label{sec:requirements}

Written requirements are not available for the autopilot and 
controls software of the TCM because it was not intended 
for embedded production-level code.
The original implementers of the software informed us that their release 
process was based on a side-by-side comparison of behaviors between 
the simulator and an experimental aircraft.  
However, for the purpose of our study, we need 
safety properties representative of those used in the certification of civil aviation transport vehicles. For this reason, NASA's Armstrong Flight Research Center chose relevant requirements from the Federal Aviation Regulations (FARs) governing commercial aviation transport vehicles (Part 25)~\cite{FAARegsPt25}, such as the following: 

\begin{quote}
\emph{FAR-25.672b}: The design of the stability augmentation system or of any other automatic or power operated system must permit initial counteraction of failures of the type specified in section 25.671(c) without requiring exceptional pilot skill or strength, by either the deactivation of the system, or a failed portion thereof, or by overriding the failure by movement of the flight controls in the normal sense.
\end{quote}

FAR requirements such as the one above refer to several high-level systems, but in our study we focus on the Guidance, Navigation, and Control (GNC) system of \figurename~\ref{fig:Controls}. The GNC is divided into three types of components:
(i) \emph{mode logic} capturing the modes in which the autopilot can operate,
how these modes are enabled and disabled, and which control systems are enabled by particular modes, 
(ii) components targeting \emph{controllability}
of the airplane by ensuring that the actuators can, at any instant, respond appropriately to a command, and (iii)
components ensuring the \emph{stability} and \emph{maneuverability} of the aircraft, making the aircraft robust to 
state disturbances and easily controlled by the pilot.

The TCM does not contain detailed  Simulink models for Navigation. Moreover, classical mathematical techniques like Lyapunov theory~\cite{Lyapunov1892} provide well-understood ways of checking stability and maneuverability properties. The formal verification tools that our study targets are better suited for checking mode logic and simple controllability properties of type (i) and (ii) above. As a consequence, we focus on GNC-level requirements, and sub-requirements (also called ``child"-requirements) that refer exclusively to Guidance. A child-requirement of the above FAR requirement, is, for example, the following:

 \begin{quote}
\emph{GNC-150}: The Guidance Navigation and Control Function shall enable the pilot to transition the vehicle from one flight condition to another (i.e. climb to level flight) under all operating conditions including failure of a single engine.
\end{quote}

In order to elicit such higher-level requirements into properties that can be checked on the components of our case-study, we examined pilot training materials for the Boeing 737 (B737) Automatic Flight Systems~\cite{SmartCockpit}.
B737 is within the class of vehicles that the TCM simulator targets. 
Within these training materials were behavioral specifications for the B737 Altitude Acquire, Altitude Hold, and Level Change Modes, equivalent in functionality to the TCM's Altitude Controller.
We also used the specifications for the B737 Heading Select Mode as the desired properties for the TCM Heading Controller, the B737 Glide Slope Capture mode for the TCM Flight Path Angle Controller, and the B737 Autothrottle for the TCM's Autothrottle.
We started with 88 GNC requirements, which we tried, with the help of these B737 documents, to map into safety properties of the Guidance system. This effort resulted into 20 properties, illustrated in Table~\ref{tab:tcm_prop}.
These 20 properties are sub-requirements of GNC-level requirements, such as the GNC-150 requirement shown above. 
In the rest of this paper, we will use property G-120 as a running example:
\begin{quote}
\emph{G-120}: The guidance shall be capable of climbing 
at a defined rate, to be limited by minimum and maximum engine 
performance and airspeeds.
\end{quote}
\begin{table*}
\caption{A summary of verified properties on the TCM.}
\label{tab:tcm_prop}
\centering
\begin{small}
\begin{tabular}{|l|l|l|l|}
\hline
\# & Property &  Assumptions  & Original Requirement\\
\hline
\cellcolor{gray!25}1& \cellcolor{gray!25}G-250 & \cellcolor{gray!25}G-260 & \cellcolor{gray!25}The heading control mode, when selected, sends roll\\
\cellcolor{gray!25}& \cellcolor{gray!25}& \cellcolor{gray!25}
&\cellcolor{gray!25}commands to turn to and maintain the commanded heading. \\
\hline
\cellcolor{gray!25}2&\cellcolor{gray!25}G-110 & \cellcolor{gray!25}G-220,G-260& \cellcolor{gray!25}The guidance system shall be capable of steering to and\\
\cellcolor{gray!25}&\cellcolor{gray!25} &\cellcolor{gray!25}&\cellcolor{gray!25}following a specified heading. \\
\hline
\cellcolor{gray!25}3&\cellcolor{gray!25}G-120 &\cellcolor{gray!25}G-180,A1,A2, & \cellcolor{gray!25}The guidance shall be capable of climbing at a defined rate, to be\\
\cellcolor{gray!25} &\cellcolor{gray!25} &\cellcolor{gray!25} FPA1&\cellcolor{gray!25}limited by minimum and maximum engine performance and airspeeds. \\
\hline
\cellcolor{gray!25}4&\cellcolor{gray!25}G-130& \cellcolor{gray!25}G-180,A1,A2 & \cellcolor{gray!25}The guidance shall be capable of descending at a defined rate, to be\\
\cellcolor{gray!25} & \cellcolor{gray!25} & \cellcolor{gray!25}&\cellcolor{gray!25}limited by minimum and maximum engine performance airspeeds. \\
\hline
\cellcolor{gray!25}5&\cellcolor{gray!25}G-140&\cellcolor{gray!25}G-120,G-200 & \cellcolor{gray!25}The guidance shall be capable of climbing at a specified rate\\
\cellcolor{gray!25} & \cellcolor{gray!25} & \cellcolor{gray!25}
&\cellcolor{gray!25}to a specified altitude, to be limited by maximum engine \\
\cellcolor{gray!25} & \cellcolor{gray!25}&\cellcolor{gray!25} &\cellcolor{gray!25}performance for a set airspeed \\  
\hline
\cellcolor{gray!25}6& \cellcolor{gray!25}G-150& \cellcolor{gray!25}G-180,A1,G-120,&\cellcolor{gray!25}The guidance shall be capable of descending at a specified rate\\
\cellcolor{gray!25} &\cellcolor{gray!25} &\cellcolor{gray!25}A2,G-200 &\cellcolor{gray!25}to a specified altitude, to be limited by maximum engine\\
\cellcolor{gray!25} &\cellcolor{gray!25} &\cellcolor{gray!25} &\cellcolor{gray!25}performance for a set airspeed \\  
\hline
\cellcolor{green!25}7 & \cellcolor{green!25}G-170 (Mode)& \cellcolor{green!25}-- &\cellcolor{green!25}The altitude control shall engage when the altitude control mode \\
\cellcolor{green!25}&\cellcolor{green!25}&\cellcolor{green!25}&\cellcolor{green!25}is selected and when the FPA control mode is not selected, and when\\
\cellcolor{green!25}&\cellcolor{green!25}&\cellcolor{green!25} &\cellcolor{green!25}there is no manual pitch or manual roll command from the stick.\\
\hline
\cellcolor{green!25}8&\cellcolor{green!25}G-180 (Mode) & \cellcolor{green!25}-- &\cellcolor{green!25}\cellcolor{green!25}The FPA control shall engage when the FPA mode \\
\cellcolor{green!25}&\cellcolor{green!25}&\cellcolor{green!25}&\cellcolor{green!25}is selected, and when there is no manual pitch or manual roll \\
\cellcolor{green!25}&\cellcolor{green!25}&\cellcolor{green!25} &\cellcolor{green!25}command from the stick.\\
\hline
\cellcolor{green!25}9&\cellcolor{green!25}G-100 & \cellcolor{green!25}-- &\cellcolor{green!25}The Guidance system shall be capable of maintaining a steady speed \\
\cellcolor{green!25}&\cellcolor{green!25}&\cellcolor{green!25}&\cellcolor{green!25}in the normal flight envelope. \\
\hline
\cellcolor{green!25}10&\cellcolor{green!25}G-200 &\cellcolor{green!25}-- &\cellcolor{green!25}If the altitude control is engaged, once the plane is within 250 ft of \\
\cellcolor{green!25}&\cellcolor{green!25}&\cellcolor{green!25}&\cellcolor{green!25}the commanded altitude, the plane will remain within 250 ft \\
\cellcolor{green!25}&\cellcolor{green!25}&\cellcolor{green!25}&\cellcolor{green!25}of the commanded altitude. \\
\hline
\cellcolor{green!25}11&\cellcolor{green!25}G-210 (Mode) & \cellcolor{green!25}-- &\cellcolor{green!25}If the FPA control and the altitude control are both selected, the FPA\\
\cellcolor{green!25}&\cellcolor{green!25}&\cellcolor{green!25}&\cellcolor{green!25}control will disengage and the altitude control will engage once the\\
\cellcolor{green!25}&\cellcolor{green!25}&\cellcolor{green!25} &\cellcolor{green!25}lane is within 200 ft of the commanded altitude.\\
\hline
\cellcolor{green!25}12&\cellcolor{green!25}G-220 (Mode) & \cellcolor{green!25}-- &\cellcolor{green!25}The heading control shall engage when the heading control mode \\
\cellcolor{green!25}&\cellcolor{green!25}&\cellcolor{green!25}&\cellcolor{green!25}is selected, and when there is no manual pitch or manual roll \\
\cellcolor{green!25}&\cellcolor{green!25}&\cellcolor{green!25}&\cellcolor{green!25}command from the stick.\\
\hline
\cellcolor{green!25}13&\cellcolor{green!25}G-230 & \cellcolor{green!25}-- &\cellcolor{green!25}If the altitude control is engaged with no active speed control, \\
\cellcolor{green!25}&\cellcolor{green!25}&\cellcolor{green!25}&\cellcolor{green!25}the speed control shall engage and the speed command shall synchronize \\
\cellcolor{green!25}&\cellcolor{green!25}&\cellcolor{green!25} &\cellcolor{green!25}to the current speed, which shall become the new altitude's target speed.\\
\hline
\cellcolor{green!25}14&\cellcolor{green!25}G-240 &\cellcolor{green!25}--&\cellcolor{green!25}The bank angle limit is established by the Bank Angle Limit Selector.\\
\hline
\cellcolor{green!25}15&\cellcolor{green!25}G-260 (Mode) &\cellcolor{green!25}--&\cellcolor{green!25}When the heading control mode is engaged, roll commands \\
\cellcolor{green!25}&\cellcolor{green!25}&\cellcolor{green!25}&\cellcolor{green!25}are given to turn in the nearest direction to the selected heading. \\
\hline
\cellcolor{green!25}16&\cellcolor{green!25}G-270 (Mode)&\cellcolor{green!25}--&\cellcolor{green!25}Manually positioning the thrust levers does not cause \\
\cellcolor{green!25}&\cellcolor{green!25}&\cellcolor{green!25}&\cellcolor{green!25}autothrottle disengagement. \\
\hline
\cellcolor{green!25}17&\cellcolor{green!25}G-290 &\cellcolor{green!25}--  &\cellcolor{green!25}The autothrottle will be limited by the max and the min throttle.\\
\hline
\cellcolor{red!25}18&\cellcolor{red!25}G-160&\cellcolor{red!25}--  
&\cellcolor{red!25}The guidance function shall be able to automatically deploy spoilers\\
\cellcolor{red!25}&\cellcolor{red!25} &\cellcolor{red!25} &\cellcolor{red!25}to limit speed in a descent, or when a significant reduction in  \\
\cellcolor{red!25} &\cellcolor{red!25} &\cellcolor{red!25} &\cellcolor{red!25}airspeed is requested by the pilot, deactivating at low speed. \\
\hline
\cellcolor{red!25}19&\cellcolor{red!25}G-280 &\cellcolor{red!25}--  
&\cellcolor{red!25}The FCCs shall issue a warning when the commanded altitude\\
\cellcolor{red!25}&\cellcolor{red!25} &\cellcolor{red!25}
&\cellcolor{red!25}disagrees with the stored commanded altitude stored in the FCCs.\\
\hline
\cellcolor{red!25}20&\cellcolor{red!25}G-190 & \cellcolor{red!25}-- &\cellcolor{red!25}If any control surface actuator loses hydraulic pressure, \\
\cellcolor{red!25}&\cellcolor{red!25} &\cellcolor{red!25} &\cellcolor{red!25}the autopilot shall disengage.\\
\hline
\end{tabular}
\end{small}
\end{table*}
\subsection{Formalization}
%
%
For verification, the properties shown in  Table~\ref{tab:tcm_prop}
must be disambiguated and formulated in terms of the signals
(inputs and outputs) of the TCM model. Moreover, they have
to be written in a formal language which, in our case, is Simulink
(see Section~\ref{sec:verification}). 
The requirements formalization process was performed 
over several iterations, and involved discussions 
between domain and verification experts. It included several steps,
presented in this section.

{\it 1. Develop a shared understanding of the requirement}. Natural
language often allows for slightly different
interpretations. 
In some cases, properties even interfered or contradicted one another.
For example, what does  
``... to be limited by minimum and maximum engine performance ...''
mean in the context of G-120? It could mean that a. the guidance shall 
be capable to climb 
when commanded to do so, but the climb rate should be limited by engine performance 
and airspeeds, or that b. the guidance is only required to be able to climb when 
the defined rate is within
the minimum and maximum engine performance.
For all such cases, we consulted with the domain expert on the 
team to develop a common
understanding and/or we refined properties to be more precise
(e.g., interpretation a. was selected for property G-120).

{\it 2. Decompose the property into a requirement on the control system
and assumptions about the physics of the airplane}. 
Since we reason formally only about controllability, 
we decomposed the properties into a part that can be 
proven formally and into corresponding assumptions
about stability and physics. Such assumptions were not
formally verified, but were confirmed by domain experts.
For example, while G-120 
specifies that the aircraft shall climb at a defined rate, 
the control system has got just sensors and actuators. Since
we do not mathematically specify the physics behavior, we do not have
a formal definition of what it means to climb at a defined rate in terms of the sensor
values---instead we define climbing solely in terms of the actuator commands. 
Since we focus on instantaneous controllability, we require that the control system 
outputs a value to an actuator that moves the aircraft into
the right direction (e.g., if the current climb rate is smaller
than the commanded climb rate, the control system should 
issue a command to the ailerons that would pitch the aircraft upwards). 
We then assume that the inner loop controllers and physics will result in an increased
climb rate. 

{\it 3. Identify affected components and signals of the control model.}
At this stage properties are still formulated in natural language
and state vague things like ``The guidance shall be capable ...''. 
Such expressions cannot be mapped to the TCM directly. 
The model describes sensors and signals. We therefore have to express 
properties in terms of the signals available in our model.
In our study, we motsly relied on domain experts to help us with this step.
In the case of G-120, we had to refine ``being capable'' as the concrete
situations in which the control system is expected to act.
This could be expressed as  ``If in FPA-control mode, and if there is no 
manual aileron or pitch command from the pilot ...'', which can be mapped to
signals in the model.

{\it 4. Decompose the requirement on the control system into 
sub-requirements on single components}. 
In some cases, the formalized requirements  
specify behavior of the complete control system in terms of its 
global inputs and outputs. 
However, proving the requirements
marked gray (7-17) in Table~\ref{tab:tcm_prop}
required information about internal signals between
lower level components. 
We therefore decomposed these requirements into 
sub-requirements over internal signals between components.
These sub-requirements were not merely slices of the global
property but actual assume/guarantee pairs that we derived
manually. 

Property G-120, for example, was decomposed as illustrated in
Fig.~\ref{fig:g120} (the figure displays Lustre code, as translated
from Simulink by our compiler).
The decomposition expresses the fact that the FPA control module,
when engaged, is in charge of maintaining an FPA (FPA1).
The mode logic ensures that the guidance system cannot
be in Altitude Control mode and FPA control mode at the same time
(i.e., these modes are mutually exclusive).  Based on this fact, the remaining
properties express that:
-- in FPA control mode, the FPA control module is engaged (G-180);
-- if not in Altitude control mode, then the Altitude control module
is not engaged (A1), and when not engaged, the Altitude control module
will not send commands (A2).
As a result, when in FPA control mode, the FPA will be the only mode
engaged, and it will issue commands to maintain an FPA, which means that
the guidance system is able to climb at a defined rate.

Property formalization helped us identify missing functionalities in 
the TCM model and also led us to refine many properties.
\begin{itemize}
\item Three requirements specify behavior for components not modeled by the TCM
(e.g., spoilers). These
requirements could not be formalized.
\item All properties that included the
behavior of the mode logic had to be made more precise. 
At this point, the requirements were precise enough
to have a unique formal representation as (temporal)
logic formulas over signals of the TCM.
\item We defined five sub-requirements 
(e.g., A1 and A2 in Fig.~\ref{fig:g120})
for properties that were verified
compositionally.
\end{itemize}
All requirements except the six for
the mode logic and two requirements on operational
limits required making assumptions about 
the physics of the aircraft.

\section{Verification of the TCM}
\label{sec:verification}

This section describes our verification efforts for the TCM: how we handle Simulink models, and how we encode and verify properties. The complete TCM benchmark (Simulink models and Lustre code) can be found in \cite{tcm_models}.

\subsection{Handling Simulink Models}
To apply SMT-based model checking, we compile the TCM Simulink model into the synchronous dataflow language Lustre~\cite{lustre1,lustre2}. In the following, we briefly introduce Lustre and then describe the compilation process and how safety properties are encoded and verified. 

\paragraph{{\bf Lustre}}
Synchronous languages are a class of languages proposed for the design of reactive systems (i.e., systems that maintain a permanent interaction
with their physical environment). Such languages are based on the theory of synchronous
time, in which the system and its environment are considered to both view time
with some ``abstract'' universal clock. Lustre combines each data stream with an associated clock as a means to
discretize time. The overall system is considered to have a universal clock
that represents the smallest time span the system is able to distinguish, together with
additional, coarser-grained, user-defined clocks. Therefore the overall system
may have different subsections that react to inputs at different frequencies. At
each clock tick, the system is considered to evaluate all streams, so all values
are considered stable for any actual time spent in the instant between ticks. Lustre programs and subprograms are expressed in terms of
\textit{Nodes}. Nodes directly model subsystems in a modular fashion, with an
externally visible set of inputs and outputs. A \textit{node} can be seen as a
mapping of a finite set of input streams (in the form of a tuple) to a finite
set of output streams (also expressed as a tuple). 

\paragraph{\bf{Simulink to Lustre}}
In Matlab/Simulink from MathWorks$^{\copyright}$\footnote{\url{http://www.mathworks.com/}}, dynamic systems are modeled as block diagrams. Simulink uses dataflow-oriented block diagram notation which consists of blocks and lines. Blocks represent either some kind of functionality, like mathematical or logical functions, or they are used for structuring the model in terms of subsystem blocks, port blocks, bus blocks etc. Every block is defined by its type and its block parameters.

 We have developed a tool called GAL\footnote{\url{https://bitbucket.org/lememta/gal}} (GeneAuto for Lustre) based on the \emph{GeneAuto}\footnote{\url{http://www.geneauto.org}} tool set. The latter is a tool for the automatic code generation of Simulink models to generate C, VHDL and Ada code. Although the development of the
generator was of primary interest, the GeneAuto project also put an emphasis on
the qualification of the toolchain~\cite{ERTS10a} by providing
traceability information all along the code generation process. 

\begin{figure*}[!t]
\centering
    \includegraphics[scale=.4]{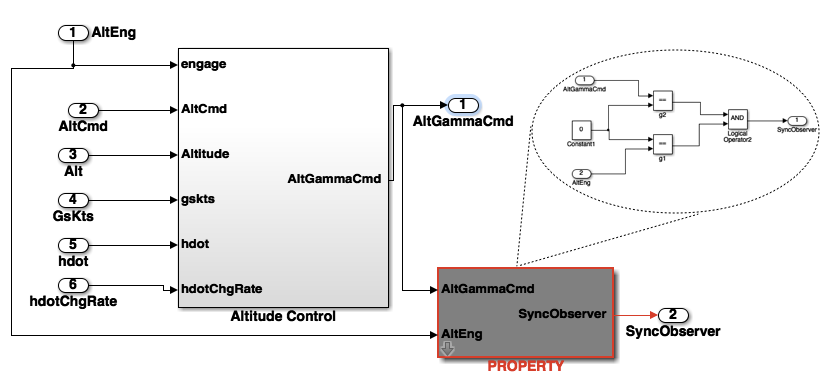}
\caption{Simulink model of the Altitude Controller subsystem with a safety property encoded as synchronous observer.}
\label{fig:annotated-simulink}
\end{figure*}

GAL can only translate a subset of Simulink blocks. This subset can be characterized as a collection of the most basic discrete-time blocks in Simulink. In a typical controller model built by control engineers, one is likely to encounter additional blocks such as the \textit{transfer function} block, the \textit{saturation} blocks, the \textit{dead-zone} block, and the \textit{integrator} block. These blocks have to first be transformed into equivalent Simulink models that GAL can handle. We have developed an automated pre-processor for Simulink models, that transforms an arbitrary Simulink model into a model ``digestible'' by GAL. The pre-processor also generates an equivalence check between the original and transformed models, modulo finite-precision arithmetic and discretization. The check can be performed using standard simulation techniques supported by Matlab. Note that our pre-processor provides automated support for transformations that are standard among aerospace engineers when using the MathWorks$^{\copyright}$ Simulink Coder\footnote{\url{http://mathworks.com/products/simulink-coder/}}.

\paragraph{{\bf Encoding safety properties}}
An extensively used technique to define expected behavior is 
\textit{synchronous observers}~\cite{DBLP:conf/amast/HalbwachsLR93}. Synchronous
observers provide an alternative to
temporal logics for specifying safety properties; 
the benefit of observers is that they
express properties in the same notation 
as the system model~\cite{Rushby:2012}.  
Observers
are typically used for simulation and testing purposes.
A synchronous observer is a wrapper used to test observable properties
of a node $N$ with minimal modification to the node itself; it returns
an error signal if the property does not hold. The task of checking
the property is thus reduced to simply checking if the stream is constantly true.

Synchronous observers are expressed in Simulink using a masked subsystem 
block. A subsystem block is a container for a set of blocks. Masking a 
block means extending it with some additional parameters. An example of such synchronous observers 
expressed in Simulink is given in Figure~\ref{fig:annotated-simulink}. The red blocks encode the safety property expressed as synchronous observer. GAL translates these blocks as a property annotation to be proven in Lustre. 

Specifically, a Lustre observer is a node taking as input all the
flows relevant to the safety property to be specified, and computing
a Boolean flow (e.g., ``Obs'' in  Fig.~\ref{fig:g120}) which is true as long as the observed
flow satisfies the property. We have used PKind~\cite{KahTin-PDMC-11} to prove the safety properties of Lustre programs. PKind is a parallel k-induction-based model checker~\cite{KahTin-PDMC-11}, which includes automated invariant generation based on templates~\cite{KahGT-NFM-11} and abstract interpretation~\cite{GarocheKT13}. 

\subsection{Safety verification results}

Table~\ref{tab:tcm_prop} summarizes the verification results of the 20 safety properties on the TCM model. At the beginning of the verification process, we discovered several modeling errors within the TCM, which led to the falsification of some properties:
\begin{enumerate}
\item \textit{Some components produced output when disabled (e.g., the altitude controller)}. 
%
%
This happened because the TCM model given to us was incomplete: the mode logic
was not implemented completely (which also affected the mode logic properties).
We remedied these problems by incorporating the necessary mode logic into the model. 
\item \textit{Manual inputs from the pilot did not override the outputs of the autopilot for all three axes}. This was again due to an incompleteness in the TCM model. We added a Simulink block (before the final output of the autopilot) to reflect the fact a pilot has the ability to override the autopilot output. 
\item \textit{Some inputs were not variables but appeared
as fixed constant values in the model (e.g., the 
bank angle limit of G-240)}. This was simply a modeling error, and it was easily corrected by modifying the appropriate variables.
\item \textit{G-180 had to be refined to 
resolve a conflict with G-210 and the 
implicit assumption that only the FPA control
or the altitude control can be active at any
moment in time.} 
\end{enumerate}

The results obtained for verification after the above changes are described in Table~\ref{tab:tcm_prop}. The properties are colored according to the verification technique used. Gray properties (7-17) are the ones proved via a compositional argument. Green properties (1-6) are the ones proved with a direct (non-compositional) proof technique. Red properties (17-20) are the ones that could not be proven on this specific model of the TCM. The latter properties applied to the B737 vehicle (see Section~\ref{sec:requirements}), however they referenced functionalities not implemented in the TCM. It took an average of 2secs for PKind to verify the green properties. 

All properties were first attempted directly, without a compositional argument. PKind was unable to verify the gray properties, despite a very high timeout setting (5 hours). Since $k$-induction is sound but not complete, this result has two possible interpretations: these properties are $k$-inductive for an extremely high $k$ or, more plausibly, they are not $k$-inductive for any $k$.  We have also tried other Lustre verification tools based on different verification techniques. Specifically, we used Kind-2\footnote{\url{http://kind2-mc.github.io/kind2/}} and Zustre\footnote{\url{www.bitbucket.org/lememta/zustre}}. Kind-2 is a complete re-implementation of PKind that also adds a verification engine based on IC3~\cite{Bradley11}, while Zustre is a tool based on the generalized property-oriented reachability implemented in Z3~\cite{pdr}. Both tools were not able to prove the gray properties.

In a second step, we decomposed the properties either in terms of component-level properties (e.g. G-120, G-130, G-140, G-150), or in ``simpler'' properties to deal with (e.g. G-250, G-110). We now give details of the compositional analysis of G-120 (see Section~\ref{sec:requirements} for the natural language description). This property involves 3 components of the TCM longitudinal control system: the \textit{Mode Logic}, the \textit{Altitude} controller and the \textit{Flight Path Angle} (FPA) controller. In order to prove G-120, we decomposed the property into 4 component-level properties: G-180 and A1 for the \textit{Mode Logic}; A2 for the \textit{Altitude} controller and FPA1 for the \textit{FPA} controller. After proving the component level properties, one still needs to make a formal compositional argument that these properties imply the system-level property G-120.  The latter argument is captured in Figure~\ref{fig:g120}. The upper box shows the Lustre nodes of the various components involved in G-120. Each component comes with its own guarantees. Such guarantees are used as assumptions in proving G-120 (described in the lower box).

\begin{figure}[!t]
\begin{lustre}
node AutoPilot(HeadMode, AilStick, ElevStick, AltMode: real; 
    FPAMode, ATMode, AltCmd, Altitude, CAS,CASCmdMCP: real;) 
returns (HeadEng, AltEng, FPAEng, ATEng: bool; CASCmd: real);
let
  -- G-180
  assert (FPAMode = 0.0) or (not (AilStick = 0.0)) or 
  	     (not (ElevStick = 0.0)) or (FPAEng = true);
  -- A1
  assert (not (AltMode= 0.0)) or (AltEng = false);
tel 

node AltitudeControl (AltEng: bool; AltCmd, Alt: real; 
			GsKts, Hdot, HdotChgRate: real) 
   returns (AltGammaCmd: real ) ;  
let
  -- A2
  assert (AltEng = true) or (AltGammaCmd = 0.0);
tel 

node FPAControl(Engage: bool; AltGammaCmd, Gamma: real; 
                ThetaDeg, VT: real) 
   returns (PitchCmd, PrePitchCmd: real);
let
 -- FPA1
 assert true -> (Engage = false) or (AltGammaCmd =  Gamma) 
 or((AltGammaCmd > Gamma) and (PitchCmd > pre(PrePitchCmd)))
 or((AltGammaCmd < Gamma) and (PitchCmd < pre(PrePitchCmd)));
tel
\end{lustre}
\begin{lustre}
node G-120 (HeadMode, AilStick, ElevStick, AltMode: real;
         FPAMode, ATMode, AltCmd, Altitude, CAS: real; 
         CASCmdMCP, Gskts, Hdot, HDotChgRate, GammaCmd: real; 
         Gamma, ThetaDeg, VT: real) 
returns (Obs: bool);
var 
  AltGammaCmd, FPain, TAlt, TFpa : real;
  HeadEng, AltEng, FPAEng, ATEng : bool;
  CasCmd, PitchCmd, PrePitch : real;
let
  HeadEng,...,CasCmd = AutoPilot(HeadMode,...,CASCmdMCP);   
  AltGammaCmd =  AltitudeControl(AltEng,...,HDotChgRate);
  PitchCmd, PrePitch = FPAControl(FPAEng,...,VT);
  
  assert FPain = (AltGammaCmd + GammaCmd);
  assert (AltMode = 0.0);
  assert (not (FPAMode = 0.0));
  assert (ElevStick = 0.0);
  assert (AilStick = 0.0);
  assert (GammaCmd > 1.0 and GammaCmd < 10.0);

  Obs = true -> (GammaCmd =  Gamma) 
      or ((GammaCmd > Gamma) and (PitchCmd > pre(prePitch)))
      or ((GammaCmd < Gamma) and (PitchCmd < pre(PrePitch)));
  --!PROPERTY: obs = true;
tel
\end{lustre}
\caption{Compositional argument for G-120 property.}
\label{fig:g120}
\end{figure}






\section{Lessons Learned}
\label{sec:lessons}

In this section, we summarize our experience and lessons learned from the application of formal methods to the TCM case study. Some findings confirmed our expectations: for example, we anticipated the fact that we would need to consult with domain experts both for requirements elicitation, and for assumption generation. We had, however, underestimated the extent to which this would be required. Others surprised us: it took an extremely long time to identify a case study that is both representative of flight-critical systems, and publicly available; the requirements elicitation and formalization phases were also much more involved than we expected. Regarding how far we would be able to go with this case study, we had no expectations to begin with, since we knew nothing about the system that we analyzed when we started our work. 

{\it 1. Case studies are hard to find}. It is difficult to obtain real case studies that are not proprietary and that can be shared outside an institution. The TCM is available for General Purpose Release from NASA Langley, with case number of LAR-18322-1. The process for obtaining the code is detailed in the latest NASA Software Catalog\footnote{\url{http://technology.nasa.gov/NASA\_Software\_Catalog\_2014.pdf}}. The TCM benchmark used in this paper can be found in~\cite{tcm_models}.

{\it 2. System description must be massaged}. Despite the progress in automating verification techniques, a huge amount of effort still needs to be placed in such a task. First of all, it is rare that verification tools are able to directly handle all the features of the languages in which systems are expressed. Despite the race towards keeping verification tools up-to-date, modeling or programming languages are typically a step ahead. There is therefore always an initial step involved, where the system description is massaged to be handled by the targeted verification tools.

{\it 3. Requirements elicitation}. Requirements of flight-critical systems are often hard to identify~\cite{aseAutoResolver,DBLP:journals/amai/GiannakopoulouBSEH11}. Even when requirements are available, they are very often written in natural language and need to be translated into a clear notation with unambiguous semantics. We had seriously under-estimated the effort that would be required in coming up with requirements that we could verify. We involved collaborators from NASA's Armstrong Flight Research Center, and still required a lot of additional effort to bridge high-level requirements with verifiable, component-level ones. It is the first time that we  had to tackle requirements starting from FARs, and we appreciate the complexity of certification tasks, and the use of safety cases to organize them~\cite{AdvoCATE} (we used safety case tools to organize our requirements, but cannot present this work here due to limited space). 

{\it 4. Incomplete requirements and assumptions}. Our case study confirmed that requirements are often incomplete or even wrong. Developers often make assumptions about the physics or the environment of a system that are not explicitly expressed, and without which  requirements do not hold. The capability to analyze requirements with automated tools is invaluable in identifying such problems with requirements. For example, the analysis we performed for property G-120 (see Section~\ref{sec:verification}) revealed the fact that an implicit assumption needed to be formalized and become part of the requirements.

{\it 5. Scalability}. The amount of progress in automating verification has been substantial. However, there will always be properties and systems on which the verification does not scale, unless more sophisticated compositional approaches are introduced to break the problem into smaller, more manageable, tasks. Compositional verification was needed to address 6 of the 20 requirements that we studied (gray properties in Figure~\ref{tab:tcm_prop}). Decomposing requirements was a non-trivial, manual task. Although our tools for this type of system do not directly support automated assumption generation and compositional verification yet, we believe that such techniques could facilitate the application of our approaches~\cite{DBLP:conf/issta/HowarGR13}.

{\it 6. Domain expertise}. Our case study proved that we are not yet at the point where formal verification can occur in the absence of domain expertise. All of the activities in observations 2, 3, and 4 above, required extensive discussion with a domain expert that was part of the team assembled for this study.

{\it 7. Verification tools}. Automated formal verification tools have made tremendous progress in recent years. They are able to cope with the growing complexities of systems. In our work, this was the key enabler in carrying out the safety analysis of TCM. Specifically, SMT-based model checking was quite effective in discharging the safety properties. In certain components of the TCM we had to deal with nonlinear arithmetic operations (e.g., trigonometric functions). While nonlinear arithmetic operations are not fully supported in current SMT solvers, we were able to cope with that by using uninterpreted functions~\footnote{The idea is to substitute nonlinear functions with uninterpreted one. This could lead to non-feasible counterexamples. In this case we add additional constraints to the uninterpreted function in order to eliminate such counterexamples.}. Handling nonlinear arithmetic operations is essential for the verification of flight-critical systems; it is therefore desirable to develop tools that can robustly handle these features.

{\it 9. Level of effort}. Verifying the TCM required approximately three person months (see Table~\ref{tab:LOE}), with the involvement of verification and aeronautics experts. Most of our time was spent eliciting the properties, formalizing them, and creating the assumptions about the physical environment that we need for verification. We also spent several weeks working towards utilizing our verification within a future safety case effort.
The actual verification process was automated, and required the least time.

\begin{table}[!ht]
\caption{TCM verification: Approximate level-of-effort}
\centering
\begin{tabular}{| c | c |}
\hline
Effort & Person Months \\
\hline
Implementation (Tools) & 0.50 \\
Preparation of Models & 0.50 \\
Property Elicitation & 0.25 \\
Formalization (Relation to model) & 0.50 \\
Physical Assumptions & 0.50 \\
Compositional Arguments & 0.25 \\
Verification & 0.20 \\
Safety Case Generation & 0.50 \\
\hline
\end{tabular}
\label{tab:LOE}
\end{table}


\section{Related Work}
\label{sec:related}


Our aim in designing this study was to make it as realistic, independent, and shareable as possible, and hence we targeted a system that is representative of flight-critical systems, that was developed outside our group, and that is available to the research community. It is hard to find realistic studies in the research community that are not proprietary and that can therefore be used as benchmarks.

In recent work, we have applied probabilistic verification and synthesis techniques to analyze the ACAS X onboard collision avoidance system~\cite{DBLP:conf/tacas/EssenG14}. Moreover, we have developed a testing infrastructure for the automated analysis of the AutoResolver air-traffic control system, aimed at the prediction and resolution of aircraft loss of separation~\cite{aseAutoResolver}. A previous large study performed by our group aimed at comparing model checking, static analysis, runtime analysis and testing, through their application for finding bugs in the Executive component of an autonomous robot developed at NASA Ames~\cite{DBLP:journals/fmsd/BratDGGHLPVVW04}. These systems are not publicly available.

Several studies related to the verification of flight-critical systems have been performed by Rockwell Collins. In \cite{DBLP:journals/cacm/MillerWC10}, the authors report on the use of their automated framework to verify Simulink and Stateflow designs of three aeronautics components: the 
ADGS-2100 Window Manager, for ensuring that data from different applications is routed to the correct aircraft display panel;  two components of the operational flight program of an unmanned aerial vehicle developed by Lockheed Martin Aerospace, one involving  redundancy management, and the other one in charge of generating actuator commands for the aircraft's six control surfaces. These studies confirmed the applicability and benefits of formal verification techniques in the design of flight-critical systems. The commercial components that were targeted are not publicly available.

More recently, Rockwell Collins has developed compositional techniques for scalable verification of architectural models expressed in the AADL language~\cite{DBLP:conf/nfm/CoferGMWLS12}. Funded through a NASA NRA, they have recently applied compositional verification to TCM architectural models, in the context of the same program milestone that drove this work\footnote{The case studies performed in this work can be found in \url{www.github.co/smaccm}}. 

The Astr\'{e}e static analyzer has been used to prove the absence of runtime errors from two Airbus components implemented in C~\cite{DBLP:dblp_conf/safecomp/SouyrisD07,BlanchetCousotEtAl02-NJ}, as well as from a C version of the automatic docking software of the Jules Vernes Automated Transfer Vehicle (ATV) enabling ESA to transport payloads to the International Space Station~\cite{conf/dasia/Bouissouetal2009}. 
Galdino et al~\cite{DBLP:conf/wollic/GaldinoMA07} use the PVS theorem prover to formally  verify an air-traffic control resolution and recovery algorithm. 
In the domain of hybrid system verification, Platzer and Clarke~\cite{DBLP:conf/fm/PlatzerC09} have applied the KeYmaera verification tool to prove properties of curved flight collision avoidance maneuvers. 
Esteve et al. have applied a probabilistic model checker to determine properties of an early design spacecraft model for the European Space Agency~\cite{DBLP:conf/icse/EsteveKNPY12}. 

For an extensive study of success stories related to the application of formal verification in practice, we refer the reader to the following technical report by Garavel and Graf~\cite{Garavel2014}. Note that, in this paper, we focus on case studies related to avionics; several other studies of safety-critical systems have been performed, for example in the contexts of medical devices and of the automotive industry.

 \section{Conclusion}
\label{sec:conclusion}

To summarize, we demonstrated a verification approach for the TCM controls system: a publicly available, realistic and complex flight-critical system of moderate size.
This study required a significant amount of effort from a team made up of both verification and domain experts.
Compositional verification was required to prove some of the safety properties of the system.
The only safety properties we did not prove in this study were those in which the desired functionality had not actually been modeled.
Our experience highlights the promise of compositional verification in the certification of flight-critical systems.
In practice, we saw that the most significant part of our effort was in defining and formalizing the appropriate properties, starting from high-level FARs requirement, all the way down to properties of the target system. We hope that our experience will be useful to other researchers and will encourage them to apply and evaluate alternative techniques on the TCM case study. To this aim, we have made all the artefacts of our benchmark publicly available.



\bibliographystyle{abbrv}
\bibliography{biblio}

\begin{thebibliography}{10}

\bibitem{tcm_models}
{TCM} models and lustre code.
\newblock Avaiable on
  \url{https://bitbucket.org/lememta/tcm_comp_verification/wiki/Home}.

\bibitem{BlanchetCousotEtAl02-NJ}
B.~Blanchet, P.~Cousot, R.~Cousot, J.~Feret, L.~Mauborgne, A.~Min{\'e},
  D.~Monniaux, and X.~Rival.
\newblock Design and implementation of a special-purpose static program
  analyzer for safety-critical real-time embedded software, invited chapter.
\newblock In T.~Mogensen, D.~Schmidt, and I.~Sudborough, editors, {\em The
  Essence of Computation: Complexity, Analysis, Transformation. Essays
  Dedicated to Neil D{.} Jones}, LNCS 2566, pages 85--108.
  Springer\discretionary{-}{}{-}Verlag, Oct. 2002.

\bibitem{conf/dasia/Bouissouetal2009}
O.~Bouissou, E.~Conquet, P.~Cousot, R.~Cousot, J.~Feret, K.~Ghorbal,
  E.~Goubault, D.~Lesens, L.~Mauborgne, A.~Min\'e, S.~Putot, X.~Rival, and
  M.~Turin.
\newblock Space software validation using abstract interpretation.
\newblock In {\em Proc. of the Int. Space System Engineering Conf., Data
  Systems in Aerospace (DASIA 2009)}, volume SP-669, pages 1--7, Istambul,
  Turkey, May 2009. ESA.

\bibitem{Bradley11}
A.~R. Bradley.
\newblock {SAT}-based model checking without unrolling.
\newblock In {\em Proceedings of the 12th International Conference on
  Verification, Model Checking, and Abstract Interpretation}, VMCAI'11, pages
  70--87. Springer-Verlag, 2011.

\bibitem{DBLP:journals/fmsd/BratDGGHLPVVW04}
G.~P. Brat, D.~Drusinsky, D.~Giannakopoulou, A.~Goldberg, K.~Havelund, M.~R.
  Lowry, C.~S. Pasareanu, A.~Venet, W.~Visser, and R.~Washington.
\newblock Experimental evaluation of verification and validation tools on
  {M}artian rover software.
\newblock {\em Formal Methods in System Design}, 25(2-3):167--198, 2004.

\bibitem{lustre1}
P.~Caspi, D.~Pilaud, N.~Halbwachs, and J.~A. Plaice.
\newblock Lustre: a declarative language for real-time programming.
\newblock In {\em Proceedings of the 14th ACM SIGACT-SIGPLAN symposium on
  Principles of programming languages}, POPL '87, pages 178--188. ACM, 1987.

\bibitem{DBLP:conf/nfm/CoferGMWLS12}
D.~D. Cofer, A.~Gacek, S.~P. Miller, M.~W. Whalen, B.~LaValley, and L.~Sha.
\newblock Compositional verification of architectural models.
\newblock In {\em {NASA} Formal Methods - 4th International Symposium, {NFM}
  2012, Norfolk,VA, USA, April 3-5, 2012. Proceedings}, pages 126--140, 2012.

\bibitem{AdvoCATE}
E.~Denney, G.~Pai, and J.~Pohl.
\newblock {AdvoCATE}: An assurance case automation toolset.
\newblock In {\em SAFECOMP Workshops}, pages 8--21, 2012.

\bibitem{DBLP:conf/icse/EsteveKNPY12}
M.~Esteve, J.~Katoen, V.~Y. Nguyen, B.~Postma, and Y.~Yushtein.
\newblock Formal correctness, safety, dependability, and performance analysis
  of a satellite.
\newblock In {\em 34th International Conference on Software Engineering, {ICSE}
  2012, June 2-9, 2012, Zurich, Switzerland}, pages 1022--1031, 2012.

\bibitem{FAARegsPt25}
{Federal Aviation Administration}.
\newblock Electronic code of federal regulations.

\bibitem{DBLP:conf/wollic/GaldinoMA07}
A.~L. Galdino, C.~M. {n}oz, and M.~{Ayala-Rinc\'{o}n}.
\newblock Formal verification of an optimal air traffic conflict resolution and
  recovery algorithm.
\newblock In {\em Logic, Language, Information and Computation, 14th
  International Workshop, WoLLIC 2007, Rio de Janeiro, Brazil, July 2-5, 2007,
  Proceedings}, pages 177--188, 2007.

\bibitem{Garavel2014}
H.~Garavel and S.~Graf.
\newblock Formal methods for safe and secure computer systems.
\newblock Technical Report BSI-Study 875, Bundesamt fuer Sicherheit in
  Informationstechnik, December 2013.

\bibitem{GarocheKT13}
P.-L. Garoche, T.~Kahsai, and C.~Tinelli.
\newblock Incremental invariant generation using logic-based automatic abstract
  transformers.
\newblock In {\em NFM}, pages 139--154, 2013.

\bibitem{DBLP:journals/amai/GiannakopoulouBSEH11}
D.~Giannakopoulou, D.~H. Bushnell, J.~Schumann, H.~Erzberger, and K.~Heere.
\newblock Formal testing for separation assurance.
\newblock {\em Ann. Math. Artif. Intell.}, 63(1):5--30, 2011.

\bibitem{aseAutoResolver}
D.~Giannakopoulou, F.~Howar, M.~Isberner, T.~Lauderdale, Z.~Rakamaric, and
  V.~Raman.
\newblock Taming test inputs for separation assurance.
\newblock In {\em 19th {IEEE/ACM} International Conference on Automated
  Software Engineering {(ASE} 2014)}, 2014.

\bibitem{lustre2}
N.~Halbwachs, P.~Caspi, P.~Raymond, and D.~Pilaud.
\newblock The synchronous dataflow programming language {L}ustre.
\newblock In {\em Proceedings of the IEEE}, pages 1305--1320, 1991.

\bibitem{DBLP:conf/amast/HalbwachsLR93}
N.~Halbwachs, F.~Lagnier, and P.~Raymond.
\newblock Synchronous observers and the verification of reactive systems.
\newblock In {\em AMAST}, pages 83--96, 1993.

\bibitem{pdr}
K.~Hoder and N.~Bj{\o}rner.
\newblock Generalized property directed reachability.
\newblock In A.~Cimatti and R.~Sebastiani, editors, {\em Theory and
  Applications of Satisfiability Testing -- SAT 2012}, volume 7317 of {\em
  LNCS}, pages 157--171. 2012.

\bibitem{DBLP:conf/issta/HowarGR13}
F.~Howar, D.~Giannakopoulou, and Z.~Rakamaric.
\newblock Hybrid learning: interface generation through static, dynamic, and
  symbolic analysis.
\newblock In {\em International Symposium on Software Testing and Analysis,
  {ISSTA} '13, Lugano, Switzerland, July 15-20, 2013}, pages 268--279, 2013.

\bibitem{Hueschen2011}
R.~M. Hueschen.
\newblock Development of the {T}ransport {C}lass {M}odel ({TCM}) aircraft
  simulation from a sub-scale {G}eneric {T}ransport {M}odel ({GTM}) simulation.
\newblock Technical report, NASA, Langley Research Center, Hampton, VA, August
  2011.

\bibitem{KahGT-NFM-11}
T.~Kahsai, Y.~Ge, and C.~Tinelli.
\newblock Instantiation-based invariant discovery.
\newblock In {\em NFM}, volume 6617 of {\em LNCS}, pages 192--207, 2011.

\bibitem{KahTin-PDMC-11}
T.~Kahsai and C.~Tinelli.
\newblock {PKIND:} a parallel $k$-induction based model checker.
\newblock In {\em PDMC}, volume~72 of {\em EPTCS}, pages 55--62, 2011.

\bibitem{Lyapunov1892}
A.~Lyapunov.
\newblock {\em General problem of the stability of motion}.
\newblock PhD thesis, Univ. Kharkov, 1892.

\bibitem{DBLP:journals/cacm/MillerWC10}
S.~P. Miller, M.~W. Whalen, and D.~D. Cofer.
\newblock Software model checking takes off.
\newblock {\em Commun. {ACM}}, 53(2):58--64, 2010.

\bibitem{DBLP:conf/fm/PlatzerC09}
A.~Platzer and E.~M. Clarke.
\newblock Formal verification of curved flight collision avoidance maneuvers: A
  case study.
\newblock In A.~Cavalcanti and D.~Dams, editors, {\em FM 2009: Formal Methods,
  16th International Symposium on Formal Methods}, volume 5850 of {\em LNCS},
  pages 547--562. Springer, 2009.

\bibitem{Rushby:2012}
J.~Rushby.
\newblock The versatile synchronous observer.
\newblock In {\em 15th Brazilian conference on Formal Methods: foundations and
  applications}, SBMF'12, pages 1--1, Berlin, Heidelberg, 2012.
  Springer-Verlag.

\bibitem{SmartCockpit}
SmartCockpit.
\newblock {B737} automatic flight systems summary.

\bibitem{DBLP:dblp_conf/safecomp/SouyrisD07}
J.~Souyris and D.~Delmas.
\newblock Experimental assessment of {Astrée} on safety-critical avionics
  software.
\newblock In {\em SAFECOMP}, pages 479--490, 2007.

\bibitem{ERTS10a}
A.~Toom, N.~Izerrouken, T.~Naks, M.~Pantel, and O.~Ssi-Yan-Kai.
\newblock Towards reliable code generation with an open tool: Evolutions of the
  {Gene-Auto} toolset.
\newblock In {\em ERTS}, http://www.sia.fr, 2010. Soci{\'e}t{\'e} des
  Ing{\'e}nieurs de l'Automobile.

\bibitem{DBLP:conf/tacas/EssenG14}
C.~von Essen and D.~Giannakopoulou.
\newblock Analyzing the next generation airborne collision avoidance system.
\newblock In {\em Tools and Algorithms for the Construction and Analysis of
  Systems - 20th International Conference, {TACAS} 2014, Held as Part of the
  European Joint Conferences on Theory and Practice of Software, {ETAPS} 2014,
  Grenoble, France, April 5-13, 2014. Proceedings}, pages 620--635, 2014.

\end{thebibliography}


\end{document}